\documentclass[superscriptaddress,aps,pra,twocolumn,preprintnumbers,amsmath,amssymb,floatfix,groupedaddress]{revtex4-1}
\usepackage[dvips]{graphicx}
\usepackage{dcolumn}
\usepackage{bm}
\usepackage{xr}
\usepackage{hyperref}
\usepackage{color}

\newcommand{\ket}[1]{\ensuremath{\left|#1\right\rangle}}

\newcommand{\Real}{\textrm{Re}}
\newcommand{\Imag}{\textrm{Im}}

\begin{document}

\title{Dispersive optical nonlinearities in an EIT-Rydberg medium}
\author{ Jovica Stanojevic, Valentina Parigi, Erwan Bimbard,
Alexei Ourjoumtsev and Philippe Grangier}
\affiliation{Laboratoire Charles Fabry, Institut d'Optique, CNRS, Univ.
Paris Sud, 2 av. Augustin Fresnel, 91127 Palaiseau cedex, France }
\begin{abstract}

We investigate dispersive optical nonlinearities that arise from Rydberg excitation blockade in cold Rydberg gases. We consider a two-photon transition scheme and study the non-linear response to a weak optical probe in presence of a strong control beam. For very low probe fields, the dominant nonlinearities are of the third order and they can be exactly evaluated in a steady state regime. In a more general case, the change in average atomic populations and coherences due to Rydberg interactions can be characterized by properly defined scaling parameters, which are generally complex numbers but in certain
situations take the usual meaning of the number of atoms in a blockade sphere. They can be used in a simple ``universal scaling" formula to determine the dispersive optical nonlinearity of the medium. We also develop a novel technique to account for the Rydberg interaction effects, by simplifying the treatment of nonlocal interaction terms, the so-called collisional integrals. We find algebraic relations that only involve two-body correlations, which can be solved numerically. All average populations and coherences are then obtained straightforwardly.
\end{abstract}
\maketitle

\section{Introduction}

The strong interactions between Rydberg atoms \cite{Gallagher1994} are studied and exploited  in various fields ranging from molecular and ultracold plasma physics \cite{Bendkowsky2009,Pohl2011} to nonlinear optics and quantum computing \cite{Pritchard2012,Saffman2010}. In particular, many efforts are currently dedicated to create large optical non-linearities in Rydberg atomic ensembles and make them observable at a few-photon level in order to use them for quantum information processing.

Using an atomic cloud as a non-linear optical medium often involves Electromagnetically Induced Transparency  (EIT) schemes \cite{Fleischhauer2005}, where the propagation of one beam (probe) is strongly modified by the presence of another beam (control), due to a two-photon resonance condition. By varying the control field, one can convert optical excitations into atomic ones, creating dark-state polaritons \cite{Fleischhauer2000} which propagate through the atomic cloud at a very slow speed. This increases the non-linear effects related to the anharmonic level structure of each individual atom \cite{Harris1990,Lukin2000,Bajcsy2009}, but so far it was not sufficient to make them observable at a few photon level. On the other hand, it has been demonstrated that when the two-photon transition involves a Rydberg state, the non-linear response of the atomic cloud can be strongly enhanced \cite{Mohapatra2008,Pritchard2010,Parigi2012,Hoffman2012,Pritchard2012}. In a simple physical picture, this happens because the interactions bring neighboring atoms out of the two-photon resonance, so each excited Rydberg atom becomes surrounded by a blockade volume where no other Rydberg excitation is possible. These ``blockade'' interactions between the polaritons increase the optical non-linearity which can become significant even at a single-photon level \cite{Gorshkov2011,Shahmoon2011}. In the resonant, dissipative regime, it has recently been demonstrated that this phenomenon could be used to operate the atomic cloud as a single-photon source \cite{Saffman2002,Dudin2012,Bariani2012,Peyronel2012,Stanojevic2012,Gorshkov2012}.

Many applications in quantum information processing require to transpose these effects from the dissipative to the dispersive regime, in order to enable coherent photon-photon interactions. Large Rydberg-induced dispersive nonlinearities have been recently measured \cite{Parigi2012} even though the regime of single-photon nonlinearities has not been reached. In this paper we study the dispersive  nonlinearities in a Rydberg EIT medium using several theoretical methods,  with experimental parameters corresponding to ref. \cite{Parigi2012}.  The results are compared to the experiment, and  the advantages and disadvantages of these various methods are discussed in detail.

\section{Theory}\label{sec:theory}

\begin{figure}
\centerline{
\includegraphics[scale = 0.25]{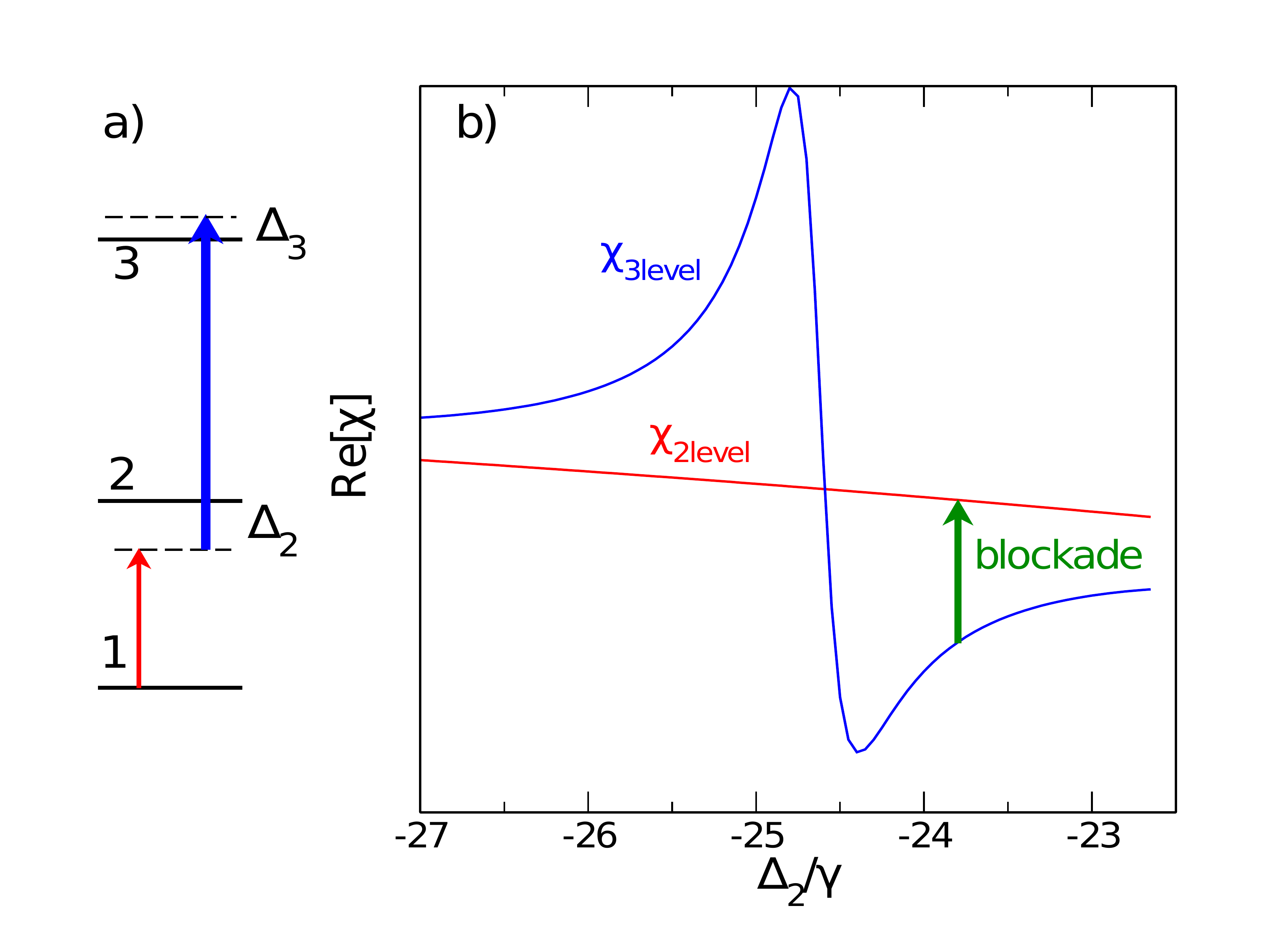}
}
\caption{
a) An ensemble of three-level atoms is excited by a strong (blue) control field, and
a weak (red) probe field. The two-photon detuning $\Delta_3$ is
much less in magnitude than the single photon detuning $\Delta_2$. 
b) Effects of Rydberg interactions in the vicinity of the two-photon resonance. Since the strong interactions effectively decouple the atoms from the control field, the blockaded atoms behave as two-level atoms which means that the resonance feature (blue) of the susceptibility of three-level non-interacting atoms gradually diminishes and moves towards the featureless two-level dependence.\label{fig:scheme}}
\end{figure}

We consider an ensemble of $N\gg 1$ three-level atoms with a ground state $\left|1\right>$, a short-lived intermediate state $\left|2\right>$, and a highly excited long-lived Rydberg state $\left|3\right>$ (see Fig. \ref{fig:scheme}). We study the non-linear response of the ensemble to a probe beam with a Rabi frequency $\Omega_p$ and optical frequency $\omega_p$, detuned from the $\ket{1}\rightarrow\ket{2}$ transition by $\Delta_2=\omega_p-\omega_{12}$, in presence of a strong control beam with a Rabi frequency $\Omega_c$ and optical frequency $\omega_c$, with a two-photon detuning $\Delta_3=\omega_p+\omega_c-\omega_{13}$.
It is assumed that the interaction between two atoms $i$ and $j$ in
a Rydberg state is described by a single potential $k_{ij}$  (as discussed in Section \ref{sec:results},
in case of multiple potential curves one may use an effective single potential).
In the rotating-wave approximation, the many-body dynamics is governed by the  Hamiltonian
\begin{eqnarray}
 H & =& \sum_{i=1}^N \bigg[-\Delta_2 \hat\sigma^i_{22} -\Delta_3  \hat\sigma^i_{33} +
\Omega^{*}_p \hat\sigma^i_{12}+ \Omega_p \hat\sigma^i_{21}+\nonumber\\
&&\hspace{8 mm}\Omega_c(\hat\sigma^i_{23}+\hat\sigma^i_{32})+\sum_{j>i=1}^N k_{ij}\hat\sigma^i_{
33} \hat\sigma^j_{33}  \bigg] ,
\end{eqnarray}
where $\hat\sigma^i_{\alpha\beta}=|\alpha_i\rangle\langle \beta_i|$, and $\Omega_c$ is taken real.  We assume that the Rydberg ensembles are locally uniform meaning that the spatial variations of any  $\hat \sigma^i_{\alpha\beta}$ and of the atomic density $\eta$ are on a significantly larger scale than
 the correlation length imposed by the interactions.
Because of this, we can drop all the indices $i$, $j$ unless keeping them is necessary for the clarity of the expressions.

To include the decay  of the intermediate state and finite laser linewidths, relaxation constants $\Gamma_{\alpha\beta}=\gamma_{\alpha\beta}-i(\Delta_{\beta}- \Delta_{\alpha})$ associated to each $\sigma_{\alpha\beta}$ are introduced. Here $\gamma_{\alpha\beta}$ are the total (positive) decay constants of the coherences, and $\Delta_1\equiv0$. The optical Bloch equations for single-atom averages $\sigma^i_{\alpha \beta}=\langle
\hat\sigma^i_{\alpha \beta}\rangle$ are
\begin{eqnarray}
&&\hspace{-7 mm}\frac{d\sigma_{12}}{dt}\!=\! -\Gamma_{12} \sigma_{12}-i[\Omega_p
(1-\sigma_{33}-2\sigma_{22}) + \Omega_c \sigma_{13}],\label{s12-eq}\\
&&\hspace{-7 mm}\frac{d\sigma_{13}}{dt}\!=\! -\Gamma_{13} \sigma_{13}-i[\Omega_c
\sigma_{12} - \Omega_p \sigma_{23} + V_{13}],\\
&&\hspace{-7 mm}\frac{d\sigma_{23}}{dt}\!=\! -\Gamma_{23} \sigma_{23}-i[\Omega_c
(\sigma_{22}-\sigma_{33}) - \Omega^{*}_p \sigma_{13} + V_{23}],\\
&&\hspace{-7 mm}\frac{d\sigma_{22}}{dt}\!=\! -\gamma_{22} \sigma_{22}\!-\!i[\Omega_p
\sigma_{21}\!-\!\Omega^{*}_p\sigma_{12}+\Omega_c
(\sigma_{23}-\sigma_{32})],\\
&&\hspace{-7mm}\frac{d\sigma_{33}}{dt}\!=\!
 -\gamma_{33} \sigma_{33}
 -i\Omega_c [\sigma_{32}-\sigma_{23}],\label{s33-eq}
\end{eqnarray}
where $V^i_{\alpha\beta}=\Sigma_{j\neq i} \,k_{ij}\langle \hat\sigma^i_{\alpha
\beta}\hat\sigma^j_{33}\rangle$ are two-atom collisional integrals.
In the following, we use the shorter notation $\sigma\!\sigma_{\alpha\beta,\mu\nu}\equiv \langle \hat\sigma^i_{\alpha
\beta}\hat\sigma^j_{\mu\nu}\rangle$ and assume the symmetry relation $\sigma\!\sigma_{\alpha\beta,\mu\nu}=\sigma\!\sigma_{\mu\nu,\alpha\beta}$. For numerical evaluations, we will express all frequencies in units of the linewidth $\gamma=\gamma_{12}=\gamma_{22}/2$ of the $\ket{1}\rightarrow\ket{2}$ transition.

For some general considerations, it may be useful to solve  Eqs. (\ref{s12-eq})-(\ref{s33-eq})
 for the averages  $\sigma_{\alpha \beta}$ assuming that the
collisional integrals are known. Obviously, in the steady state such solutions are linear
combinations of the collisional integrals
  $V_{13}$, $V_{31}$, $V_{23}$, and $V_{32}$
\begin{equation}\label{s_mu-nu}
 \sigma_{\mu\nu}=\sigma^{3lev}_{\mu\nu}+\beta^{\mu\nu}_{13}V_{13}+\beta^{\mu\nu}_{31}V_{31}+
 \beta^{\mu\nu}_{23}V_{23}+\beta^{\mu\nu}_{32}V_{32},
\end{equation}
where $\beta^{\mu\nu}$ are functions of the Rabi frequencies and relaxation constants
and $ \sigma_{\mu\nu}^{3lev}$ are $ \sigma_{\mu\nu}$ in the absence of interactions.
In practice,
one would need to know these integrals explicitly. This would require
to calculate atomic correlators, which is difficult because
 $n$-body correlators depend on $(n+1)$-body correlators.

There are various approaches how to close the system of equations for
the correlators.  One possible way  is to expand the correlators in
powers of the probe Rabi frequency $\Omega_p$ \cite{Sevincli2011PRL}.
Here, we are mostly interested in the first two terms in the expansion of
$\sigma_{12}$ and  $\sigma_{33}$
\begin{eqnarray}
\sigma_{12}&=&\Omega_p\sum_{s=0}\sigma^{(2s+1)}_{12} |\Omega_p|^{2s},\\
\sigma_{33}&=&\sum_{s=1}\sigma^{(2s)}_{33}|\Omega_p|^{2s}\,.
\end{eqnarray}
Similarly, all other averages can be factorized in a leading term in $\Omega_p$, $\Omega_p^*$ and power series in $|\Omega_p|^2$.
It is easy to conclude that  $V_{13}\sim |\Omega_p|^3$ and
$V_{23}\sim |\Omega_p|^4$. Therefore,  $V_{23}$ and $V_{32}$ do not contribute to $\sigma^{(3)}_{12}$.  We will use
the propagation equation for $\langle\hat\sigma^i_{13}\hat\sigma^j_{33}\rangle$
to illustrate how the system of equations to calculate  $\sigma^{(3)}_{12}$ gets closed
\begin{eqnarray}\label{ss1333}
 \frac{d\langle\hat\sigma^i_{13}\hat\sigma^j_{33}\rangle}{dt} &\!=\!&
-(\Gamma_{13}+\gamma_{33})\langle\hat\sigma^i_{13}\hat\sigma^j_{33}\rangle
-i\bigg[\Omega_p \langle\hat\sigma^i_{23}\hat\sigma^j_{33}\rangle + \nonumber\\
&&\hspace{-15 mm} \Omega_c (\langle\hat\sigma^i_{13}\hat\sigma^j_{32}\rangle-
\langle\hat\sigma^i_{13}\hat\sigma^j_{23}\rangle)+\sum_{s\neq
i}k_{is}\langle\hat\sigma^i_{13}\hat\sigma^s_{33}\hat\sigma^j_{33}\rangle \bigg].
\end{eqnarray}
Note that all terms in the sum of three-body averages are proportional to
$|\Omega_p|^5$ in the lowest order except the term corresponding to $s=j$, which is proportional to
$|\Omega_p|^3$. Therefore, only this $s=j$ term contributes to the
third-order nonlinearities. In a steady-state, the $|\Omega_p|^3$ terms in Eq. (\ref{ss1333})
satisfy the following equation
\begin{equation}
(\Gamma_{13}\!+\!\gamma_{33}\!+i\,k_{ij})\sigma\!\sigma^{(3)}_{13,33}=\!-
i\Omega_c(\sigma\!\sigma^{(3)}_{13,32}-\sigma\!\sigma^{(3)}_{13,23}).
\end{equation}
We see that this equation does not contain any three-body terms.
Similarly, the equations for the other correlators
$\sigma\!\sigma_{\mu\nu,\alpha\beta}^{(3)}$ also have no three-body terms  and thus
$\sigma^{(3)}_{12}$ is fully determined by the expansions of the two-body correlators and single atom averages. Even though there are many two-body correlators, it turns out that we do not need to handle all of them at once because several smaller independent subsets of equations are obtained after making the expansion. In Appendix \ref{AppendixChi3}, we give more details
on the  calculation of two body correlations using the expansion in powers of $\Omega_p$.

For stronger excitation, many expansion terms are needed in order to get convergent results so this approach is not convenient anymore.  In this case, it is better to know approximately the whole sum than to know exactly a few expansion terms. Our method presented in Appendix \ref{AppendixCollInt}
estimates the collisional integrals without making use of the expansion. Because of this, it can give reasonable, albeit approximate, results in a larger range of $\Omega_p$ than a truncated expansion with a few first exact terms (some examples are given in Section \ref{sec:results}).

In more details, the calculation proceeds as follows.
The equations contain various local correlators and nonlocal collisional integrals, that are a consequence of interactions which couple different atoms and make the calculation a difficult many-body problem. In our method we first slightly modify the
 ladder approximation  \cite{Sevincli2011JPB} which is commonly applied to three-body collisional integrals in order to substitute them with two-body ones.
The small modification is that just before applying the ladder approximation we split the  collisional integrals involving three atoms in two parts corresponding to two different leading dependences in power of $\Omega_p$
\begin{eqnarray}\label{ladder-approx}
&&\hspace{-10 mm}\sum_{s\neq i}k_{is}\langle\hat\sigma^i_{\alpha3}\hat\sigma^s_{33}\hat\sigma^j_{\mu\nu}\rangle =
\delta_{3\mu}k_{ij}\langle\hat\sigma^i_{\alpha3}\hat\sigma^j_{3\nu}\rangle +\nonumber\\
&&\hspace{-4 mm}\sum_{s\neq i,j}k_{is}\langle\hat\sigma^i_{\alpha3}\hat\sigma^s_{33}\hat\sigma^j_{\mu\nu}\rangle
\approx \delta_{3\mu} k_{ij}\langle\hat\sigma^i_{\alpha3}\hat\sigma^j_{3\nu}\rangle+V_{\alpha3}\sigma^j_{\mu\nu}.
\end{eqnarray}
One may argue that
$(V_{\alpha3}-k_{ij}\sigma\!\sigma_{\alpha3,33})\sigma_{\mu\nu}$ should be used in the last equation instead
of $V_{\alpha3}\sigma_{\mu\nu}$, but the difference between these two expressions is insignificant for small $\sigma_{33}$.
By conjugating this equation one gets a similar expression for the correlators  $\sigma\!\sigma_{3\alpha,\mu\nu}$.
The approximation (\ref{ladder-approx}) is  exact up to the lowest order in powers of $\Omega_p$ in which the interaction dependence starts to appear. For instance, it is the third-order for $\sigma\!\sigma_{13,33}$ but the fourth order for $\sigma\!\sigma_{23,33}$ and $\sigma\!\sigma_{33,33}$.

In the second step, we obtain an algebraic system that only contains the collisional integrals of interest, as shown
in Appendix \ref{AppendixCollInt}.
Since we only need four $V_{\mu\nu}$ integrals to determine all $\sigma_{\alpha\beta}$, we are ultimately solving four second-order polynomials in $V_{\mu\nu}$. This is certainly much simpler than calculating a large number of coupled many-body equations.
After solving this system for four $V_{\mu\nu}$, one can get all single-atom averages using the solution (\ref{s_mu-nu})
 of Eqs (\ref{s12-eq})-(\ref{s33-eq}).

\section{Scaling of blockaded single-atom averages}\label{sec:scaling}

In this section we establish some relationships between two ways of describing the effects of interactions in Rydberg ensembles. The first one is more
 intuitive and it is based on the concept of excitation-blockade while the other approach is more formal and it uses collisional integrals $V_{\alpha\beta}$  to evaluate the effects of interactions.

The concept of an excitation blockade sphere is very useful to get an intuitive physical picture of the effects of Rydberg-Rydberg interactions on  $\sigma_{12}$. In this picture, each Rydberg atom prevents $n_b$ surrounding atoms from being excited into Rydberg states and thus these $n_b$ atoms behave effectively as two-level atoms. Consequently,  $\sigma^i_{12}$ of these blockaded atoms is equal to $\sigma_{12}^{\rm 2lev}$ of two-level atoms. On the other hand,  $\sigma^i_{12}$ of unblockaded atoms is just $\sigma_{12}^{\rm 3lev}$ of noninteracting atoms. Therefore,  $\sigma_{12}$ in an ensemble of $N$ atoms of which $N_r$ are Rydberg atoms, is a weighted average  of $\sigma_{12}^{3lev}$ for  $(N-N_r n_b)$ unblockaded atoms, and $\sigma_{12}^{2lev}$  for  $N_r n_b$  blockaded ones~:
\begin{eqnarray}\label{s12_scaling}
\sigma_{12}&=&
 \sigma_{12}^{3lev} (1-\frac{N_r n_b}{N}) + \sigma_{12}^{2lev} \frac{N_r n_b}{N}\\ \nonumber
 &=&
\sigma_{12}^{3lev}+p_r n_b(\sigma_{12}^{2lev}-\sigma_{12}^{3lev}) ,
\end{eqnarray}
where $p_r=N_r/N=\sigma_{33}$.
 Essentially the same averaging was also used in \cite{Tanasittikosol2011} to describe interaction effects in  microwave dressed Rydberg atoms. In \cite{Tanasittikosol2011}, the radius of the blockade sphere was defined by the condition that the Van der Waals shift is equal to the two-photon resonance  width, and was later used to calculate $n_b$ as the number of atoms in a blockade sphere. Also, $p_r$ was replaced by the Rydberg excitation probability $p_3$ of noninteracting atoms, which is acceptable if the blockade spheres do not overlap. 
 Another form of universal scaling supported by Monte-Carlo simulations is presented in \cite{Ates2011}.

Here, we want to use Eq. (\ref{s12_scaling}) as a definition of $n_b$ because we can calculate $\sigma_{12}$ in the presence of interactions for low probe-light  using the rate equations. From this perspective, $n_b$ is generally a complex number. Similarly to Eq. (\ref{s12_scaling}), it is easy to define a mean value of any  $\sigma_{\mu\nu}$ for $\mu,\nu\neq3$ but extending it to  $\sigma_{\mu3}$ and $\sigma_{3\mu}$ is a bit less straightforward. However, one may expect a  similar result for  $\sigma_{33}=p_r\approx p_3(1-n_b p_3)$ which just means that  $n_b p_3 N$ atoms are not available for Rydberg excitation. Obviously, the $n_b$ corresponding to $\sigma_{33}$ has to be real.
Therefore, it is also interesting to determine to what extent the same parameter $n_b$ can characterize the interaction effects on  $\sigma_{12}$  and $\sigma_{33}$.

The change in $\sigma_{12}$ induced by interactions is $\sigma^{coll}_{12}=\sigma_{12}-\sigma_{12}^{3lev}$.
To derive $n_b$ in  Eq. (\ref{s12_scaling}), we start with the solutions (\ref{s_mu-nu}) which are now functions of the collisional integrals $V_{\alpha\beta}$. We take the low probe-light limit of these solutions since we know exactly the collisional integrals in this case. Instead of using the exact expression (\ref{ss1333_general}) for the $\sigma\!\sigma_{\mu\nu,\alpha\beta}$ correlator, which is too complicated for the theoretical analysis in this section,  we use the much simpler relation (\ref{ss1333approx}),  which is still a very accurate approximation in the dispersive regime.

We assume that the probe field is sufficiently low so that the lowest correlation orders correctly describe the effects of interactions.
Using the solutions  Eq. (\ref{s_mu-nu}) for $\sigma_{12}$ and expanding it in powers of $\Omega_p$, we find the relation
\begin{equation}\label{s12_V13}
\sigma^{coll}_{12}\approx \frac{-\Omega_c\Omega_p|\Omega_p|^{2}}{\Gamma_{13}\Gamma_{12}+\Omega_c^2}V_{13}^{(3)},
\end{equation}
where  $V_{13}^{(3)}$ is the lowest expansion coefficient of
 $V_{13}=\Omega_p\Sigma_{s=1}V^{(2s+1)}_{13}|\Omega_p|^{2s}$.
According to Eqs.  (\ref{ss1333approx}), $V_{13}^{(3)}$ is
\begin{equation}\label{W13}
V_{13}^{(3)}\approx \sigma^{(1)}_{13}\sigma^{(2)}_{33} \; T  \; I_b
\end{equation}
where the spatial integral $I_b$ depends on the atomic density $\eta$, on the Rydberg potential $k_{ij}$ and on the effective relaxation constant $T\approx\Gamma_{13}+\Omega_c^2/\Gamma_{12}$ of the Rydberg transition, light-shifted and power-broadened by the strong control beam:
\begin{equation}\label{n_b-def}
I_b=\eta\int d^3 r \frac{i k_{ij}(r)}{T+i k_{ij}(r)}. 
\end{equation}
Combining Eqs. (\ref{s12_V13}) and (\ref{W13}) we get
\begin{equation}\label{s12final}
\sigma^{coll}_{12} \!=\!  p_3 I_b \frac{ i \Omega_c \sigma^{(1)}_{13}\Omega_pT  }{\Gamma_{13}\Gamma_{12}+\Omega_c^2}
\!=\! p_3 I_b \frac{i \Omega_p \Omega_c \sigma^{(1)}_{13}}{\Gamma_{12}},
\end{equation}
where we used $p_3\approx |\Omega_p|^{2}\sigma^{(2)}_{33}$.
On the other hand, the leading term of $\sigma_{12}^{2lev}-\sigma_{12}^{3lev}$ for low probe light is
\begin{equation}\label{s12diff}
\sigma_{12}^{2lev}-\sigma_{12}^{3lev} \approx  \frac{i \Omega_p\Omega_c \sigma^{(1)}_{13}}{\Gamma_{12}}
\end{equation}
Direct inspection of Eqs. (\ref{s12_scaling}), (\ref{s12final}), and (\ref{s12diff}) yields the expected identification
$n_b = I_b$,
and for the Van der Waals interaction $k=-C_6/r^6$ we find
\begin{equation}\label{n_b-complex}
n_b = I_b = \frac{ 2\pi^2\eta }{3\sqrt{iT/C_6}}.
\end{equation}
Strictly speaking, $n_b$ is a complex number and can even be dominantly imaginary. However, using the atomic cloud as as dispersive optical medium puts constraints on the signs of $\Delta_2$, $\Delta_3$ and $C_6$. In order to avoid absorption, one must keep all atoms far-detuned from all resonances. One consequence of this is that the power-boadened linewidth $\Real(T)\approx\gamma_{13}+\Omega_c^2/\Delta_2^2$ of the Rydberg transition must remain smaller than its light-shifted effective detuning $-\Imag(T)\approx\Delta_3-\Omega_c^2/\Delta_2$: for strong control beams, this requires $\Delta_2$ and $\Delta_3$ to have opposite signs. In addition, this effective detuning must have the same sign as $C_6$, otherwise some atomic pairs will be resonantly excited into Rydberg states and the optical losses will strongly increase. As a consequence, in the dispersive regime $iT/C_6$ is dominantly real and positive, so that according to the relation (\ref{n_b-complex}) $n_b$ is dominantly real and positive as well:
\begin{equation}\label{n_b-real}
n_b \approx \frac{2\pi^2\eta }{3\sqrt{\left(\Delta_3-\frac{\Omega_c^2}{\Delta_2}\right)/C_6}}.
\end{equation}
In contrast, if the effective two-photon detuning and $C_6$ have opposite signs, $n_b$ is dominantly imaginary. Using eq. (\ref{n_b-real}) and previous ones, one recovers the formulae given in the Appendix of ref. \cite{Parigi2012}.

From an experimental point of view, it is also worth noticing that in ref. \cite{Parigi2012}, the control beam is actually a standing wave, so that the control Rabi frequency is spanning the whole range from zero to its maximum value. In order to keep the above conditions valid in all points in space, it is not possible to stay on the two photon resonance as considered in ref. \cite{Sevincli2011PRL}, and one has to use a  small  but non-zero value of $\Delta_3$.

Now we want to find out how $\sigma_{33}$ is affected by interactions in the low probe-field limit. However, since  $\sigma_{33}^{(4)}$ is the lowest term  affected by interactions, all four collisional integrals contribute to
$\sigma^{coll}_{33}=\sigma_{33}-\sigma^{3lev}_{33}$.  In this case we can define the scaling parameter ${\tilde n}_b$ as follows
\begin{equation}\label{s33_scaling}
{\tilde n}_b=\frac{\sigma_{33}^{3lev}-\sigma_{33}}{\sigma_{33}^2}.
\end{equation}
This ${\tilde n}_b$ is real and has to be positive in the blockade regime so clearly ${\tilde n}_b$ and   $n_b$ cannot be exactly the same. We can compare these two scaling parameters only in the dispersive regime, where $n_b$ is also dominantly real and positive. Even though one formally needs to calculate the forth-order nonlinearities to account for the effects of interactions on $\sigma_{33}^{(4)}$, this nonlinearity is only due to binary interactions, which means that our  method in Appendix \ref{AppendixCollInt}  calculates $V^{(4)}_{23}$ exactly. By expanding Eq. (\ref{s_mu-nu}) for $\sigma_{33}$, we find that the contributions from $V_{13}$ and $V_{31}$ to $\sigma^{coll}_{33}$ are approximately
\begin{equation}\label{s33_fourth_13}
\sigma^{coll}_{33}\sim  \Real\left[\frac{\Gamma_{12}\Gamma^*_{23} \Omega^{*}_p}
{\gamma_{23}\Omega_c(\Gamma_{12}\Gamma_{13}+\Omega_c^2)} iV_{13}\right]
\end{equation}
and the contributions from $V_{23}$ and $V_{32}$ to $\sigma^{coll}_{33}$ are
\begin{equation}\label{s33_fourth_23}
\sigma^{coll}_{33}\sim  {\rm Re}\left[\frac{\Gamma^*_{23}V_{23}}
{\gamma_{23}\Omega_c} \right].
\end{equation}
Our calculation of $V_{23}$ for very low probe laser intensity shows that the  contributions from $V_{23}$ and $V_{32}$
are much smaller than those from $V_{13}$ and $V_{31}$. Therefore, both  $n_b$ and ${\tilde n}_b$ depend only on $V_{13}$ ($V_{31}$) in  the low probe-light limit. In this limit we can use the simple relation
 $V_{13}=i\Omega_p\Omega_c p_3 n_b/\Gamma_{12}$ in the expression (\ref{s33_fourth_13}) to get the following relation
\begin{equation}\label{n_b-for-s33}
{\tilde n}_b=\xi_1\Real[n_b] +\xi_2 {\rm Im}[n_b]
\end{equation}
One can verify that
 $\xi_1\approx1$ but $\xi_2$ is slightly larger than 6 for the parameters of Ref.\cite{Parigi2012} with $\Delta_3/\gamma=1/3$. This means that the difference between $\Real[n_b]$ and  ${\tilde n}_b$  can be rather large unless ${\rm Im}[n_b]$ is very small. Hence one can write $\Real[n_b] \sim {n}_b \sim {\tilde n}_b$ only for large enough two-photon detuning, e.g.  $\Delta_3/\gamma=2$, as will be also apparent  from the numerical results shown on Fig. \ref{fig:nb}.

\section{Numerical results}\label{sec:results}

In this section we present our numerical results in the dispersive regime. We use physical parameters very similar to the peak values in experiment \cite{Parigi2012} performed with an ensemble of  $^{87}$Rb atoms: $\eta=0.04$ $\mu$m${}^{-3}$, $\Delta_2/\gamma=-25$, $\gamma_{13}/\gamma=1/10$, and $\gamma/2\pi=3$ MHz. The single-photon detuning $\Delta_2$ was chosen to be negative in order to avoid heating the atoms with the probe beam, which means that dispersive effets could be observed for positive two-photon detunings $\Delta_3>0$ and attractive Rydberg states $C_6>0$ (see section \ref{sec:scaling}). Therefore, we focus on attractive Rydberg states $\ket{3}=\ket{nD_{5/2},m_J=5/2}$ with principal quantum numbers $n=46$, $50$, $56$, and $61$ as used in Ref.\cite{Parigi2012}, where the numerical values of $C_6/\gamma$  are  respectively $2400$, $5000$, $15000$, and $36000$ $\mu$m${}^{6}$. Assuming that the control laser irradiance remains constant, we take $\Omega_c/\gamma=3$ for the state $n=50$ and use the $n^{-3/2}$ dependence of the transition dipole to calculate $\Omega_c$ for the other states.

We make two additional assumptions which simplify the calculations compared to the actual situation in Ref.\cite{Parigi2012} and still provide a good qualitative agreement with the experimental data. First, we disregard the (longitudinal and transverse) spatial and temporal variations of the Rabi frequencies and of the atomic density, and consider them as static and uniform: taking these variations into account is not a trivial problem and the quantitative modeling of the experimental results will be  presented elsewhere. Second, we assume like we did before that
the interactions between Rydberg atoms are described by a single potential $k_{ij}$, which is actually only true for $S$ Rydberg states. In our case 
the $nD_{5/2}$ states with the maximal projection $m_j = 5/2$ were targeted. This implies that all ungerade
$nD_{5/2}+nD_{5/2}$ potential curves of Rb should be considered \cite{Stanojevic2006}. All these potentials are attractive but very different in strength, ranging over two orders of magnitude.
Since we are interested in the dispersive regime where none of these potentials becomes resonant, and since we consider only global properties obtained after radial and angular integration over the atomic cloud, we can try to reproduce their overall effect with a single effective potential \cite{Reinhard2007,Walker2008}.
How this potential is defined in detail depends on the problem and on the purpose for which it is used, and our procedure will be explained in detail in a separate publication. It turns out that the values for our effective $C_6^{\rm eff}$ differ by less than $20\%$ from those in \cite{Reinhard2007}. The difference arises because we average $\sqrt{C_6}$ over the relevant potentials to get an effective $\sqrt{C_6^{\rm eff}}$, which is the quantity of interest in our problem, while in \cite{Reinhard2007} $C_6$ was averaged to get  $C_6^{\rm eff}$.

The optical response of the cloud to the probe beam is characterized by its optical susceptibility $\chi_{12}$, proportional to $\sigma_{12}/\Omega_p$. By using appropriate normalization this proportionality constant can be disregarded, and in the following we will call ``susceptibility'' the ratio $\sigma_{12}/\Omega_p$. We can define the following normalized susceptibility $S_{\rm norm}$ as a measure of nonlinear effects induced by the Rydberg interactions
\begin{equation}\label{S_norm-def}
S_{\rm norm}=\frac{\sigma_{12}-\sigma^{2lev}_{12}}{\sigma^{3lev}_{12}-\sigma^{2lev}_{12}} \,.
\end{equation}
In the absence of interactions, $\sigma_{12}=\sigma^{3lev}_{12}$ holds so that $S_{\rm norm}=1$. In a fully blockaded system $\sigma_{12}\approx\sigma^{2lev}_{12}$ so that $S_{\rm norm}$ vanishes. If the atoms are placed in an optical cavity, all the coherences in Eq. (\ref{S_norm-def}) should be substituted by their projections  onto the cavity mode(s). In this case the nonlinear phase shifts are mapped into the shifts $\Phi$ of the cavity resonance frequencies, so it is useful to consider the normalized real part of the susceptibility
\begin{equation}\label{S-def}
S=\frac{{\rm Re}[\chi_{12}-\chi^{2lev}_{12}]}{{\rm Re}[\chi^{3lev}_{12}-\chi^{2lev}_{12}]} \,.
\end{equation}
because, in a single-mode cavity, it is equal to the normalized cavity resonance shift  $\Phi(|\Omega_p|^2)/\Phi(0)$  which has been measured \cite{Parigi2012}.

\begin{figure}
\includegraphics[scale = 0.34]{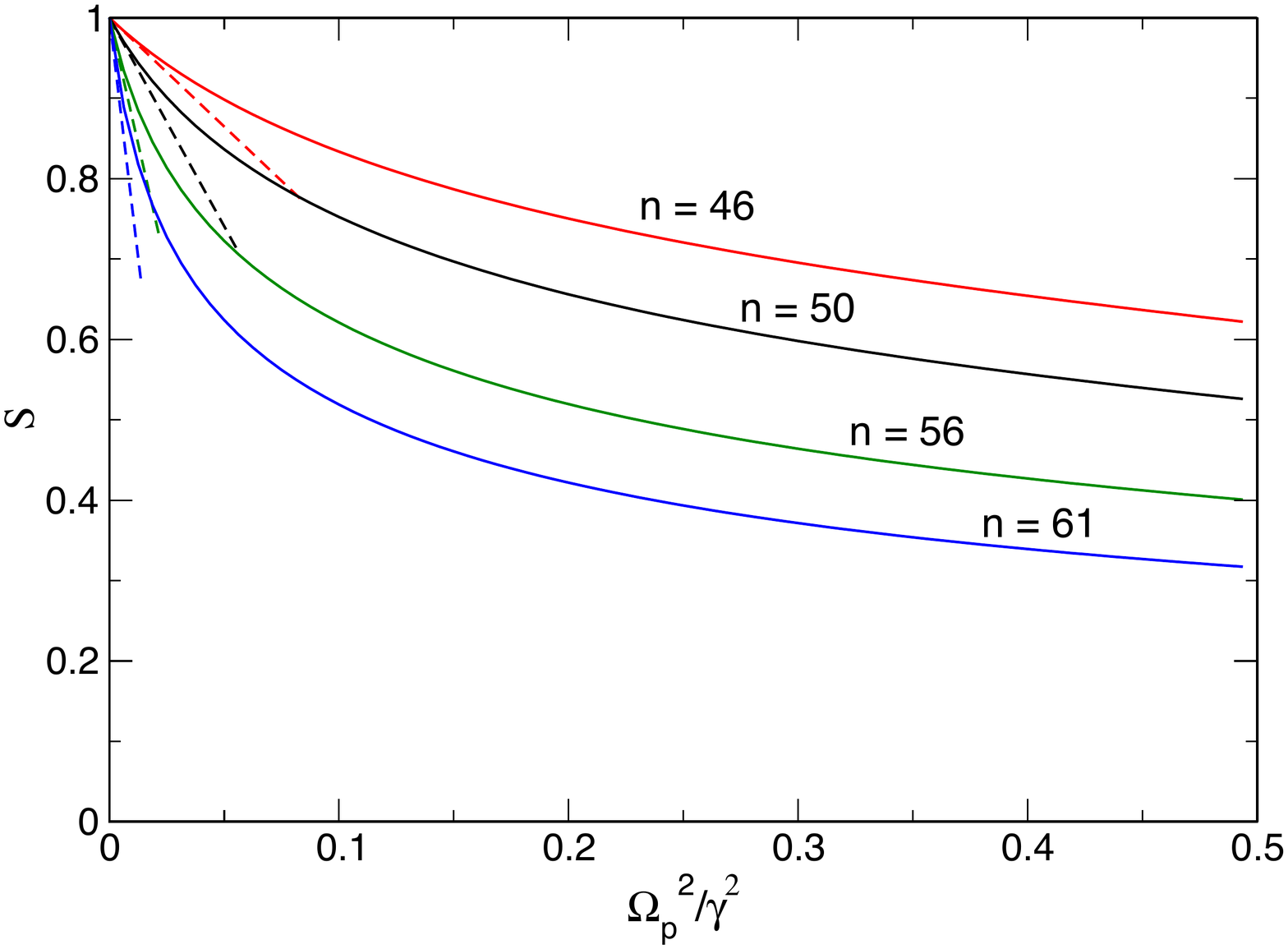}
\caption{Normalized susceptibility (\ref{S-def})
for $n=46$, 50, 56, and 61. For noninteracting  systems  $S=1$ holds everywhere while $S=0$ is reached in  the fully blockaded regime.
The slopes (dashed lines) at the origin are given by the third-order interaction-induced nonlinearities and they follow the $\sqrt{C_6}\propto n^{11/2}$ scaling as a function of the principal quantum number $n$. The solid curves are calculated using the method in Appendix \ref{AppendixCollInt}.
\label{fig:Phi}}
\end{figure}

On Fig. \ref{fig:Phi}, the full curves present the shift $S$ calculated using the methods described in Appendix \ref{AppendixCollInt}, whereas their initial slopes (dashed curves) are determined by the third-order nonlinearities evaluated using the exact expression (\ref{ss1333_general}) for $\sigma\!\sigma^{(3)}_{13,33}$, which scales as $\sqrt{|C_6|}\sim n^{11/2}$ because
$$
\left. \frac{dS}{d(|\Omega_p|^2)}\right|_{\Omega_p\rightarrow0}\!\sim\! V^{(3)}_{13}\!\sim\! n_b\!\sim\!\sqrt{|C_6|}.
$$
As pointed out in Appendix \ref{AppendixCollInt}, our approximate method should reproduce the third-order nonlinearities  exactly and this is manifestly true in Fig. \ref{fig:Phi}. Since the actual  normalized susceptibilities rapidly deviate from the initial third-order dependence, all the more as $n$ is large, this figure also shows that the higher order ($>3$) nonlinearities are important and, in fact, quickly become dominant in our parameter range. Even though all the curves show signs of saturation, they have not yet reached  the fully blockaded regime ($S=0$), which is also confirmed by the experimental results \cite{Parigi2012}: the normalized susceptibilities are getting saturated before reaching the fully blockaded regime and then slowly approach the limit $S=0$.

\begin{figure}
\includegraphics[scale = 0.30]{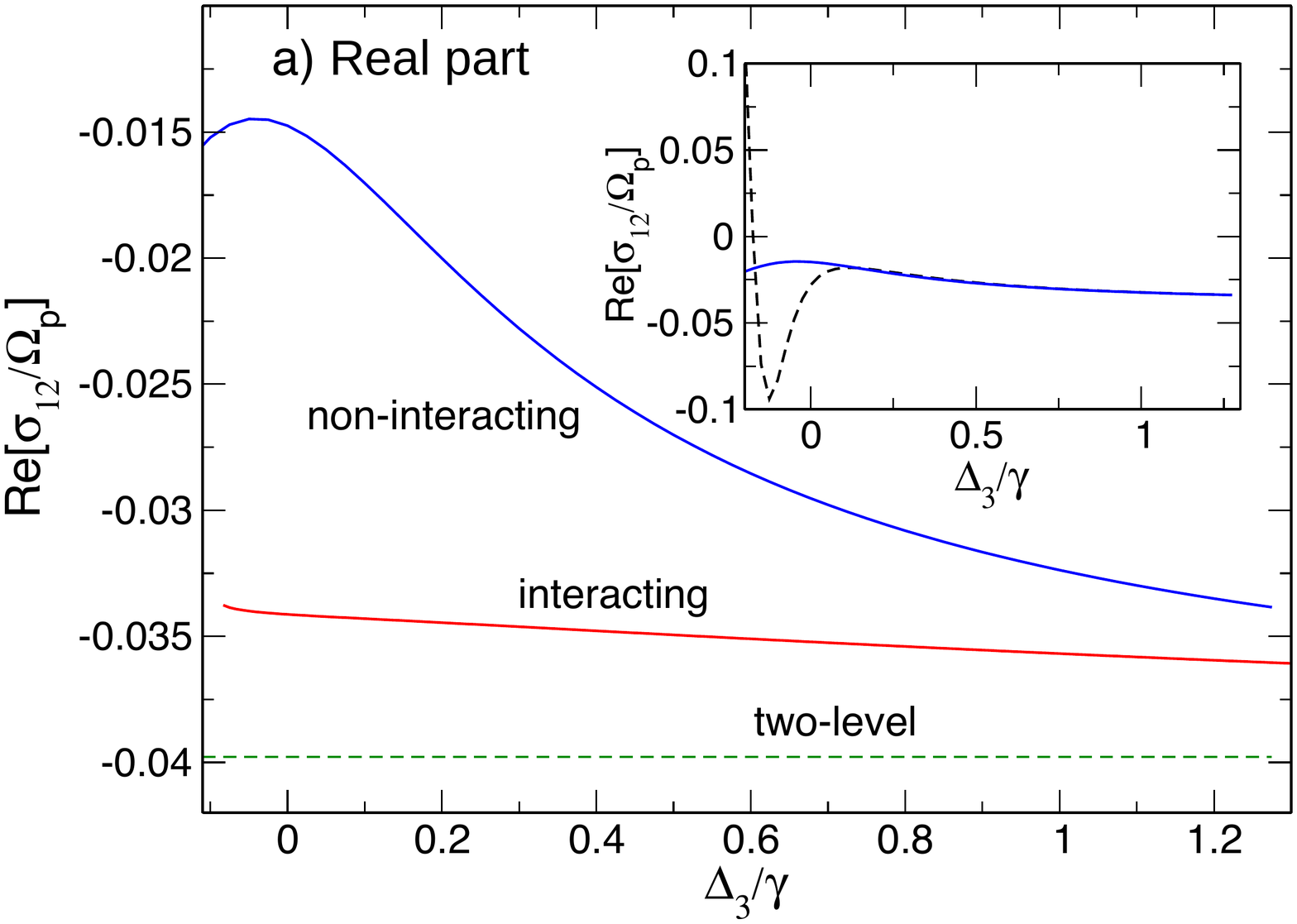}
\includegraphics[scale = 0.30]{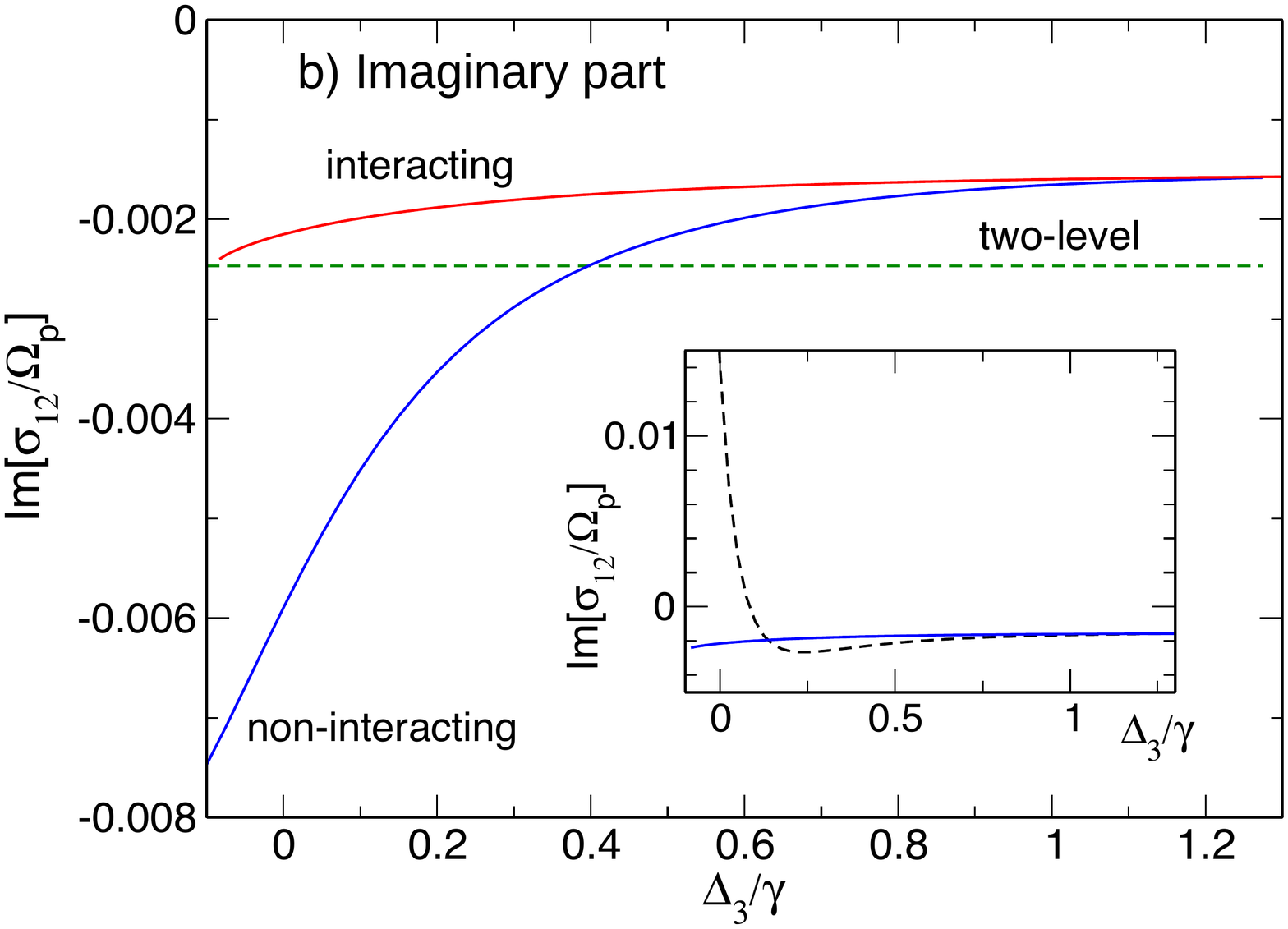}
\caption{Real and imaginary part of susceptibility $\sigma_{12}/\Omega_p$ in the vicinity of the two-photon resonance. All parameters are the same as in Fig. \ref{fig:Phi} for $n=61$,
 except that $\Omega_p$ is fixed:  $|\Omega_p|^2=\gamma^2/2$. The susceptibilities in the noninteracting (blue) and interacting case (red) are shown together with the susceptibility of two-level atoms (dashed green). In the inset plots, the noninteracting $\sigma^{3lev}_{12}/\Omega_p$(blue) is compared with the truncated expansion $\sigma^{(1)}_{12}+|\Omega_p|^2\sigma^{(3)}_{12}$ (dashed black).
\label{fig:chi}}
\end{figure}

In Fig. \ref{fig:chi} we show the real and imaginary part of  $\sigma_{12}/\Omega_p$ in the vicinity of the two photon resonance  for $n=61$. The physical parameters are  the same as in Fig. \ref{fig:Phi} for $n=61$, except that we take the maximal $|\Omega_p|=\gamma/\sqrt{2}$ considered in Fig. \ref{fig:Phi}.
To compare these results with the two limits of $S_{\rm norm}$, we show $\sigma_{12}/\Omega_p$  for non-interacting and interacting atoms together with  $\sigma_{12}/\Omega_p$ of two-level atoms. Even though the strongest interaction effects in Fig.  \ref{fig:Phi} are seen for $n=61$,  the corresponding  $\sigma_{12}$ has not yet reached the fully blockaded limit  $\sigma^{2lev}_{12}$.
As $\Delta_3$ approaches zero, the interaction and blockade effects become more pronounced.
In the inset plots, we show the noninteracting $\sigma^{3lev}_{12}/\Omega_p$ and its truncated expansion
 $\sigma^{(1)}_{12}+|\Omega_p|^2\sigma^{(3)}_{12}$ corresponding to the third-order approximation of $\sigma_{12}$. We see that this third-order approximation completely fails for negative detunings even for noninteracting atoms. For $\Delta_3/\gamma=1/3$, which was used in Fig. \ref{fig:Phi}, it remains satisfactory in the noninteracting case, but in the presence of interactions the range of $\Omega_p$ where this expansion remains valid is dramatically reduced. This confirms the result of Fig.  \ref{fig:Phi}, where the actual curves quickly deviate from their initial slopes given by the third-order non-linearity.
\begin{figure}
\centerline{
\includegraphics[scale = 0.34]{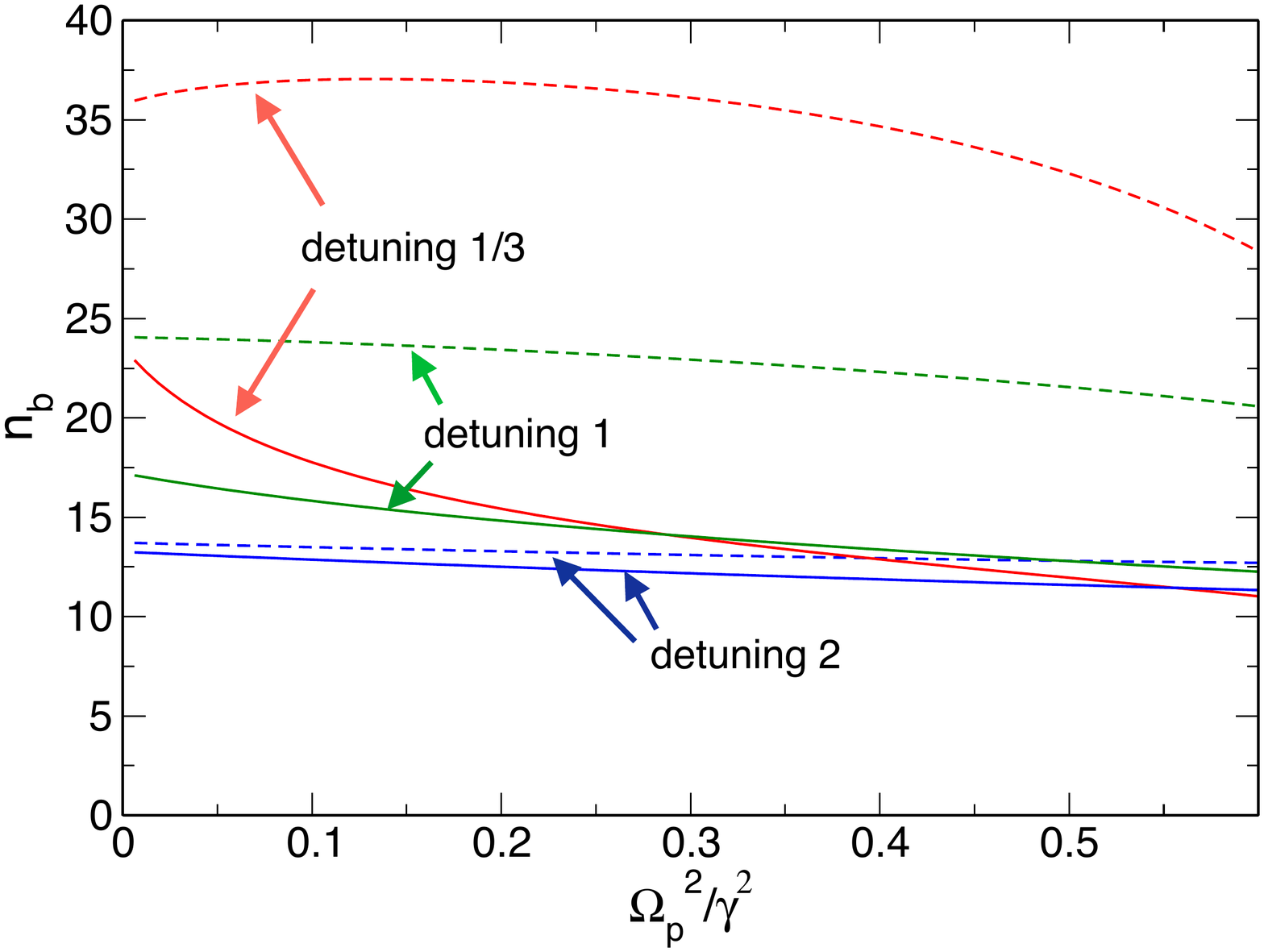}
}
\caption{Scaling parameters $\Real[n_b]$ (solid) and ${\tilde n}_b$ (dashed) for $\Delta_3/\gamma=1/3$ (red), $\Delta_3/\gamma=1$ (green), and $\Delta_3/\gamma=2$ (blue) as functions of probe-laser intensity. All the other parameters are the same as in Fig.  \ref{fig:Phi} for $n=50$.
\label{fig:nb}}
\end{figure}
The blockade spheres are expected to be ``stiff" but not really hard balls. Therefore, it is expected that $n_b$ slowly decreases as $\Omega_p$ increases.  To take this into account, we can use Eqs. (\ref{s12_scaling}) and (\ref{s33_scaling}) as definitions of $n_b$ and ${\tilde n}_b$ for any  $\Omega_p$.
Since $n_b$ is complex and ${\tilde n}_b$ is positive, we can only compare them if they are both positive.  In Fig.  \ref{fig:nb}, we show how ${\rm Re}[n_b]$ and ${\tilde n}_b$
depend on the probe-laser intensity for $\Delta_3/\gamma=1/3$, 1, and 2. All the other parameters are the same as in Fig.  \ref{fig:Phi} for $n=50$. The results demonstrate that as  $\Delta_3$ increases the imaginary part of $n_b$ vanishes, and both scaling parameters and their difference decrease as expected. One can also see a significant difference between ${\rm Re}[n_b]$ and ${\tilde n}_b$ for $\Delta_3=\gamma/3$, which is due to the rather large factor $\xi_2=6.4$, and to the non-negligible   (although small) ${\rm Im}[n_b]$. As for the contributions of  $V_{23}$ and $V_{32}$ to  ${\tilde n}_b$, we verified that they remain below $\sim3$\%.

\section{Conclusions}

In this work we investigated strong dispersive optical nonlinearities in ensembles of Rydberg atoms under EIT conditions.
Using the expansion in powers of $\Omega_p$, we evaluated the steady-state third-order susceptibility, dominant for sufficiently low probe fields. In this regime we established precise relationships between an intuitive description based on Rydberg blockade and a more formal one based on the rate equations and collisional integrals. We introduced scaling parameters $n_b$ that encapsulate the effects of interactions on single-atom average populations and coherences. Generally speaking, the $n_b$ that correspond to coherences are complex and can even become dominantly imaginary for resonant excitation of atomic pairs. However, when the Rydberg excitation blockade is used in the dispersive, off-resonant regime, we found that these scaling parameters become positive and very similar to each other, taking the usual interpretation of the number of atoms in a blockade sphere.

Beyond the lowest-order regime, like in any many-body calculation, the complexity arises from the treatment of nonlocal interaction-dependent quantities. We resolved this problem by approximating many-body correlations by two-body ones, and finding a closed set of algebraic equations satisfied by the latter. This method reproduces exactly the lowest-order interaction-induced nonlinearities, allowing us to determine the range of probe laser intensities where the lowest-order expansion is valid. As expected, this range decreases with increasing Rydberg interactions. We calculated the optical response of the atomic cloud under conditions similar to those in a recent experiment \cite{Parigi2012}, and we justified the theoretical approach that was used to interpret the results of the experiment. The numerical results are in very good qualitative agreement with the experimental data. A full quantitative analysis requires to take into account the spatial and temporal variations of Rabi frequencies; this will be the object of another work.

\vskip 0.3cm

\noindent {\it Acknowledgements.} This work is supported by the ERC  Grant 246669 ``DELPHI''.  We thank Thomas Pohl and Etienne Brion for discussions.

\appendix
\section{Third-order solution}
\label{AppendixChi3}

Here we show the main steps which yield the exact solution for the third-order susceptibilities in the presence of interactions.
Because this  solution is too long to be  explicitly written, we will just provide its general form. It turns out that the rather large system for two-body correlators decouples into several independent subsets which can be solved order by order. We are only interested in the correlators that are relevant for the collisional integral $V_{13}$. Therefore, we end up solving the following system (for $\gamma_{33}=0$)
 \begin{eqnarray}
&&\hspace{-6 mm}(i\Gamma_{13}-k_{ij})\sigma\!\sigma^{(3)}_{13,33}\!=\! \Omega_c [\sigma\!\sigma^{(3)}_{12,33} -
\sigma\!\sigma^{(3)}_{13,23} + \sigma\!\sigma^{(3)}_{13,32}],\label{correl-expand1}\\
&&\hspace{-6 mm} (i\Gamma_{13}+i\Gamma_{23}-k_{ij})\sigma\!\sigma^{(3)}_{13,23}=
-\sigma\!\sigma^{(2)}_{13,13}   +\nonumber\\
&&\hspace{21 mm} +\Omega_c \big[\sigma\!\sigma^{(3)}_{12,23}+ \sigma\!\sigma^{(3)}_{13,22}-
\sigma\!\sigma^{(3)}_{13,33}\big]\label{correl-expand2},
\end{eqnarray}
\begin{eqnarray}
&&\hspace{-6 mm}i(\Gamma^*_{23}+\Gamma_{13})\sigma\!\sigma^{(3)}_{13,32}= \sigma\!\sigma^{(2)}_{13,31}+ \Omega_c \big[\sigma\!\sigma^{(3)}_{12,32} -\nonumber\\
&&\hspace{43 mm}\sigma\!\sigma^{(3)}_{13,22} + \sigma\!\sigma^{(3)}_{13,33}\big],\\
&&\hspace{-6 mm}i\Gamma_{12}\,\sigma\!\sigma^{(3)}_{12,33}= \sigma^{(2)}_{33}\!+\! \Omega_c \big[\sigma\!\sigma^{(3)}_{13,33}+\sigma\!\sigma^{(3)}_{12,32} - \sigma\!\sigma^{(3)}_{12,23}\big],\\
&&\hspace{-6 mm} i(\Gamma_{12}+\Gamma_{23})\sigma\!\sigma^{(3)}_{12,23}=\sigma^{(2)}_{23} -\sigma\!\sigma^{(2)}_{13,12}+\nonumber\\
&&\hspace{23 mm} \Omega_c \big[\sigma\!\sigma^{(3)}_{12,22} -
\sigma\!\sigma^{(3)}_{12,33} + \sigma\!\sigma^{(3)}_{13,23}\big],\\
&&\hspace{-6 mm} i(\Gamma_{13}+\gamma_{22})\sigma\!\sigma^{(3)}_{13,22}=\sigma\!\sigma^{(2)}_{13,21} -\sigma\!\sigma^{(2)}_{13,12}+\nonumber\\
&&\hspace{23 mm} \Omega_c \big[\sigma\!\sigma^{(3)}_{12,22} +
\sigma\!\sigma^{(3)}_{13,23} - \sigma\!\sigma^{(3)}_{13,32}\big],\\
&&\hspace{-6 mm} i(\Gamma^*_{23}+\Gamma_{12})\sigma\!\sigma^{(3)}_{12,32}=\sigma\!\sigma^{(2)}_{31,12} +\sigma^{(2)}_{32}+\nonumber\\
&&\hspace{23 mm} \Omega_c \big[\sigma\!\sigma^{(3)}_{12,33} +
\sigma\!\sigma^{(3)}_{13,32} - \sigma\!\sigma^{(3)}_{12,22}\big],\\
&&\hspace{-6 mm} i(\Gamma_{12}+\gamma_{22})\sigma\!\sigma^{(3)}_{12,22}=\sigma\!\sigma^{(2)}_{12,21} -\sigma\!\sigma^{(2)}_{12,12}+\sigma^{(2)}_{22}+\nonumber\\
 &&\hspace{23 mm}\Omega_c \big[\sigma\!\sigma^{(3)}_{12,23} +
\sigma\!\sigma^{(3)}_{13,22} - \sigma\!\sigma^{(3)}_{12,32}\big].\label{correl-expand8}
\end{eqnarray}
All expansion terms are solved order by order so it is assumed that the first- and second-order expansion terms have been solved before solving the third-order terms. Therefore, all second order averages in Eqs. (\ref{correl-expand1})-(\ref{correl-expand8}) are considered to be known functions. Only two correlators $\sigma\!\sigma^{(3)}_{13,33}$   and   $\sigma\!\sigma^{(3)}_{13,23}$
 in the first two equations of the system (\ref{correl-expand1})-(\ref{correl-expand8}) are related to interaction terms
 $k_{ij} \sigma\!\sigma^{(3)}_{\alpha\beta,\mu\nu}$.
 We can eliminate all the other correlators in Eqs. (\ref{correl-expand1})-(\ref{correl-expand8}) to get
\begin{eqnarray}
 &&\hspace{-6 mm}(F_1\!-\!k_{ij})\sigma\!\sigma^{(3)}_{13,33}-K_1\sigma\!\sigma^{(3)}_{13,23}= \frac{A_1 k_{ij}}{P+k_{ij} Q}+G_1,\label{elimin_eq1}\\
 &&\hspace{-6 mm}(F_2\!-\!k_{ij})\sigma\!\sigma^{(3)}_{13,23}-K_2\sigma\!\sigma^{(3)}_{13,33}= \frac{A_2 k_{ij}}{P+k_{ij} Q}+G_2,\label{elimin_eq2}
\end{eqnarray}
where all $F_\alpha$, $K_\alpha$, $G_\alpha$, P and Q are lengthy but known functions of control Rabi frequency $\Omega_c$ and relaxation constants $\Gamma_{\mu\nu}$, $\gamma_{\mu\nu}$. From eqs. (\ref{elimin_eq1},\ref{elimin_eq2}), 
 we can find a general expression for the radial dependence of $\sigma\!\sigma^{(3)}_{13,33}$:
\begin{eqnarray} 
&&\sigma\!\sigma^{(3)}_{13,33}= \label{ss1333_general} \\
&&\frac{f_0({\bm \Gamma},{\bm \Omega})\!+\!f_1({\bm \Gamma},{\bm \Omega})k_{ij}\!+\!f_2({\bm \Gamma},{\bm \Omega})k_{ij}^2  }
{h_0({\bm \Gamma},{\bm \Omega})\!+\!h_1({\bm \Gamma},{\bm \Omega})k_{ij}\!+\!h_2({\bm \Gamma},{\bm \Omega})k_{ij}^2\!+\!h_3({\bm \Gamma},{\bm \Omega})k_{ij}^3  },\nonumber
\end{eqnarray}
where $f_i$ and $h_i$ are functions of relaxation constants and Rabi frequencies.
This general form can be analytically integrated to get the collisional integral $V^{(3)}_{13}$ which is important for the interaction induced optical nonlinearities and $\sigma\!\sigma^{(3)}_{12,33}$.

We use the exact $V^{(3)}$ based on Eq. (\ref{ss1333_general}) when we want to precisely know its numerical values, such as the slopes in Fig.  \ref{fig:Phi}.  However,
for some theoretical analysis, it may be more convenient to use  approximate but simpler expressions.
 Since we mainly consider the dispersive regime here, we can assume that $|\Delta_2|$ is much larger than any other frequency.
For very large interatomic separations $R_{ij}$ there should be no correlations between particles so that $\sigma\sigma_{\alpha\beta,\mu\nu}= \langle\sigma^i_{\alpha
\beta}\sigma^j_{\mu\nu}\rangle\rightarrow \langle\sigma^i_{\alpha
\beta}\rangle\langle\sigma^j_{\mu\nu}\rangle$ at $R_{ij}\rightarrow\infty$. For finite separations $R_{ij}$ and $|\Gamma_{12}|\gg \Omega_c$ the correlator is
\begin{equation}\label{ss1333approx}
\sigma\sigma_{13,33}^{(3)}\approx \sigma^{(1)}_{13}\sigma^{(2)}_{33}\frac{T}{T+i k_{ij}} ,
\end{equation}
where
$T\approx\Gamma_{13}+ \Omega_c^2 / \Gamma_{12}$ 
is the effective damping rate of the Rydberg transition. This is in fact an excellent approximation for $|\Gamma_{12}|\gg \Omega_c$.

\section{Collisional integrals}
\label{AppendixCollInt}

In order to evaluate the single-atom averages in the presence of interactions, we need to know the collisional
integrals  $V^i_{\alpha3}=\Sigma_{j\neq i} \,k_{ij}\langle\sigma^i_{\alpha3}\sigma^j_{33}\rangle$.
There are several well known difficulties of doing this. The first one is that $(n+1)$-body collisional integrals are needed to  evaluate $n$-body correlators. We use a slightly modified ladder approximation (\ref{ladder-approx}). The ladder approximation is a standard approach to this problem.   Another problem is that the steady-state equations mix local correlators, as functions of interatomic separations between involved atoms, and nonlocal collisional integrals. Our goal in this appendix is to explain how one can get a closed algebraic system that contains only two-body collisional integrals, which is easy to solve numerically.

We start with the rate equations for two-body correlators $\sigma\!\sigma_{\alpha\beta,\mu\nu}$ assuming that the approximation (\ref{ladder-approx}) has been utilized. Most of these equations do not explicitly contain the terms $k_{ij}\sigma\!\sigma_{\alpha\beta,\mu\nu}$, where $k_{ij}=-C_s/R_{ij}^s$. (From now on we use $k$ and $R$ instead of $k_{ij}$ and $R_{ij}$.)  By $P_{m}$, we denote  two-body correlators $\sigma\!\sigma_{\alpha\beta,\mu\nu}$ whose time derivative contain the term  $k\,P_{m}$  and $Q_{n}$ are all other two-body correlators. As mentioned previously, we assume the symmetry $\sigma\!\sigma_{\alpha\beta,\mu\nu}=\sigma\!\sigma_{\mu\nu,\alpha\beta}$. In addition, all substitutions
 \begin{equation}\label{ss11ab}
\sigma\!\sigma_{11,\mu\nu}=\sigma_{\mu\nu}-\sigma\!\sigma_{22,\mu\nu}-\sigma\!\sigma_{33,\mu\nu}
\end{equation}
are made. This results in ten equations for ten $P_{m}$ correlators and twenty six  equations for twenty six $Q_{n}$ ones which are conveniently written as follows
\begin{eqnarray}
k P_{m}&=&\sum_q a_{mq}P_q+\sum_{s} b_{ms}Q_s+R_m, \label{P_m}\\
0&=&\sum_s c_{ns}Q_s+\sum_{q} d_{nq}P_{q}+\Re_n, \label{Q_m}
\end{eqnarray}
where the coefficients $a_{mq}$, $b_{ms}$, $c_{ns}$,  $d_{nq}$ are either the Rabi frequencies or relaxation constants and  $R_m$ and $\Re_n$ contain terms which are essentially constant on the length scale of interaction induced correlations. These $R_m$ and $\Re_n$ are in general linear combinations of $\sigma_{\mu\nu}$ (originating from the substitution (\ref{ss11ab})) and terms $\sigma_{\mu\nu}V_{\alpha3}$ and $\sigma_{\mu\nu}V_{3\alpha}$ (originating from the ladder approximation (\ref{ladder-approx})). According to Eq. (\ref{s_mu-nu}), all $\sigma_{\mu\nu}$ are linear combinations of four collisional integrals $V_{13}$, $V_{31}$, $V_{23}$, and $V_{32}$.
Therefore, $R_m$ and $\Re_n$ are at most second order polynomials of  the four collisional integrals.
To additionally clarify those $P$($Q$)-correlators we give a few examples. For instance, $\sigma\!\sigma_{13,33}$ is a $P$-correlator   while $\sigma\!\sigma_{22,33}$ is a $Q$-correlator whose corresponding $\Re_n$ is equal to zero.  On the other hand, $\sigma\!\sigma_{22,13}$ is a $Q$-correlator with nonzero  $\Re_n$.

Using Eqs. (\ref{Q_m}), we can eliminate all $Q_s$ from Eqs. (\ref{P_m}) as follows

\begin{eqnarray}
&&\hspace{-11 mm} k P_{m}=\sum_q (a_{mq}+\sum_{n}\alpha_{mn}d_{nq})P_q+\\
&&\hspace{-4 mm} \sum_{s}(b_{ms}+\sum_{n}\alpha_{mn}c_{ns})Q_s+R_m +\sum_{n}\alpha_{mn}\Re_n \,.
\end{eqnarray}

Imposing the condition
\begin{equation}\label{Q_s-elimin}
b_{ms}+\sum_{n}\alpha_{mn}c_{ns}=0
\end{equation}
 for each $m$ and $s$, we get
\begin{equation}\label{P_m-new}
k P_{m}=\sum_q {\tilde a}_{mq}P_q+{\tilde R}_m ,
\end{equation}
where ${\tilde a}_{mq}=a_{mq}+\Sigma_{n}\alpha_{mn}d_{nq}$.
We can introduce matrices $\hat a$, $\hat b$,  $\hat c$, $\hat d$, $\hat \alpha$ as matrices corresponding to
the coefficients $a_{mq}$, $b_{ms}$, $c_{ns}$,  $d_{nq}$, and $\alpha_{mn}$,  respectively. From Eq. (\ref{Q_s-elimin}) we obtained $\hat \alpha =-{\hat b}{\hat c}^{-1}$. The matrix form of Eq. (\ref{P_m-new}) is
\begin{equation}\label{P-matrix}
k {\bf P}=({\hat a}-{\hat b}{\hat c}^{-1}{\hat d}){\bf P}+{\tilde{\bf R}},
\end{equation}
For any eigenvector  $\bm \lambda$ and corresponding eigenvalue $\lambda$ of the matrix $({\hat a}-{\hat b}{\hat c}^{-1}{\hat d})^T$, relation (\ref{P-matrix}) implies
\begin{equation}
k\, {\bm \lambda}.{\bf P}=\lambda \,{\bm \lambda}.{\bf P}+{\bm \lambda}.{\tilde{\bf R}},
\end{equation}
and consequently
\begin{equation}\label{P-solved}
{\bm \lambda}.{\bf P}=\frac {{\bm \lambda}.{\tilde{\bf R} }} {k-\lambda}  \,.
\end{equation}
Note that ${\bm A}.{\bf B}$ is not a scalar product but just the sum $\Sigma_m A_m B_m$.
From Eq. (\ref{P-solved}), we see that each eigenvector $\bm \lambda$ imposes one condition on the collisional integrals
$$
{\bf V}=\eta\int d^3 R \, k\,{\bf P}
$$
as follows
\begin{equation}\label{lambdaV}
{\bm \lambda}.{\bf V}=\left[\eta \int d^3 R \, \frac {k} {k-\lambda}\right]  {\bm \lambda}.{\tilde{\bf R} }  \,
=F(\lambda)\,  {\bm \lambda}.{\tilde{\bf R} }  \,.
\end{equation}
Introducing matrices
\begin{equation}
\hat U=\left[  \begin{array}{c}
  			     {\bm \lambda}_1\\
  					\vdots\\
  					{\bm \lambda}_{10}
 				 \end{array}\right],\quad
\hat F=\left[  \begin{array}{c}
  			    F(\lambda_1) {\bm \lambda}_1\\
  					\vdots\\
  					F(\lambda_{10}){\bm \lambda}_{10}
 				 \end{array}\right],\quad				
 \end{equation}

we get
$$
{\bf V}={\hat U}^{-1}{\hat F}{\tilde {\bf R} }.
$$

For the Van der Waals interaction $k=-C_6/R^6$ we find
$
F(\lambda)= (2\pi^2\eta/3) \sqrt{C_6/\lambda}.
$

We do not need to consider all ten collisional integrals ${\bf V}$ but only four $V_{13}$, $V_{31}$, $V_{23}$, and $V_{32}$ since any component of $\tilde {\bf R}$ only depends on these four collisional integrals.
Once these integrals are determined, the single-atom averages are obtained using
the solution (\ref{s_mu-nu}) of  Eqs. (\ref{s12-eq})-(\ref{s33-eq}).

\bibliography{chi3dd}

\end{document}